\documentstyle[aps,preprint,epsf]{revtex}
\begin{document} 
\draft
\title{A New Class of Resonances at the Edge of the Two Dimensional
 Electron Gas}
\author{N.B. Zhitenev, M. Brodsky, R.C. Ashoori}
\address{Department of Physics, Massachusetts Institute of Technology, Cambridge, 
Massachusetts, 
02139}
\author{M.R. Melloch}
\address{Department of Electrical Engineering, Purdue University, West Lafayette, 
Indiana  47907}
\maketitle
\date{Recieved}
\begin{abstract}
We measure the frequency dependent capacitance of a gate covering the edge and 
part of 
a two-dimensional electron gas in the quantum Hall regime. In applying a positive 
gate 
bias, we create a metallic 'puddle' under the gate surrounded by an insulating 
region. 
Charging of the puddle occurs via electron tunneling from a metallic edge 
channel. 
Analysis of the data allows direct extraction of this tunneling conductance. Novel 
conductance resonances appear as a function of gate bias. Samples with gates 
ranging 
from 1-170~$\mu$m along the edge display strikingly similar resonance spectra. 
The 
data suggest the existence of unexpected structure, homogeneous over long length 
scales, at the sample edge.
\end{abstract}

\narrowtext

Edge states of the quantum Hall effect can be thought of as ideal channels that 
carry a net current only in one direction. Under conditions of the quantum Hall 
effect 
(QHE), the bulk of a two-dimensional electron gas (2DEG) is insulating, and 
gapless 
excitations appear only near the sample edges. The conductance associated with 
an edge 
state resulting from spin-split Landau level is precisely $e^2/h$.\cite{halp} This 
result is 
true {\it independent} of the interactions between electrons and the details of the 
self-
consistent potential which forms at the edge. However, interactions between 
electrons 
may lead to novel quantum phenomena reflected in the microscopic structure of 
the of the 
electron gas near the sample edges.

In a model including a self-consistent confining 
potential,\cite{chklov,mceuen} the 
edge region breaks up into alternating sets of conducting strips which are the 
compressible 
EC and narrower insulating (rather incompressible) strips. This picture 
appears to be in reasonable qualitative agreement with recent 
experiments\cite{zhiten,takaok,hwang} under conditions of the integer QHE. 
However,
it may miss important modifications that result from 
exchange\cite{dempsey,chamon} and correlations.\cite{brey}  More recent 
theoretical 
investigations have developed a significantly different picture when electron 
correlations 
are included. New effects such as spontaneous spin 
polarization\cite{dempsey} and edge reconstructions which develop abruptly 
with 
increasing magnetic field have been predicted.\cite{chamon,fertig}

While transport {\it along} the edge is rather insensitive to edge structure, 
transport {\it across} the edge is quite sensitive to it. Strong variations in this 
conductance 
arise from the presence of the incompressible strips (IS) separating EC 
associated with 
different Landau levels and the EC from the bulk. Prior investigations of the 
resistance of 
the edge incompressible strips were based mostly on the study of the non-
equilibrium 
quasi-dc transport in 2DEG (for a review see Ref.~\cite{been}). 

In this letter, we report an experiment which isolates the 2DEG edges and 
yields a 
powerful probe of their structure. A metallic island forms when the local 
2DEG density is 
enhanced by adjusting the voltage on a gate covering the edge and part of a 
2DEG. This 
island is spatially separated from a conducting edge and the bulk by a narrow 
($\sim$0.1~$\mu$m wide)
insulating strip. Using frequency dependent capacitance measurements, 
we probe equilibrium electron tunneling across the narrow quantum Hall 
insulator. 
This technique allows us to accurately measure resistances of the strip in the 
range of 
$10^6-10^{11}\Omega$. As the gate bias is varied, we observe a series of 
sharp 
resonances in 
the tunneling conductance. This resonant structure is qualitatively 
indistinguishable for 
samples created using 1~$\mu$m or 170~$\mu$m long gates. We contend that 
these 
resonances arise from abrupt transitions in structure in or near edge 
incompressible strips. 
The sharpness of the resonances in the case of long gates indicates an unusual 
ubiquity of 
the transitions, despite the expected inhomogeneity of the sample edge.

Five samples in the form of Hall bars defined by wet etching on a Al$_{1-
x}$Ga$_x$As/GaAs wafer with mobility $\mu$=700,000~cm$^2$/Vs and 
concentration 
$n_b=1.0\times 10^{11}$~cm$^{-2}$ were used for the measurements. The 
metal gate 
covering the edge of the Hall bar and part of the sample is depicted in the inset 
to Fig.~1. 
The gate dimensions $l\times w$  (where $l$ is the length along the edge and 
$w$ is the 
width) were $170\times 7$, $32\times 5$, $3\times 5$, $3\times 3$, and 
$1\times 3$ 
$\mu m$ for samples \#1 through \#5 respectively.

To measure the small capacitance between the 2DEG and the gate 
(geometrical 
values 3fF-1pF) in the frequency range $10^2-10^7$~Hz, an ac bridge 
scheme\cite{ashoori} is implemented. The input of a high electron mobility 
transistor is 
connected to the sample's gate which forms the balancing point of the bridge. 
The circuit 
element consisting of the impedance between the 2DEG and the gate forms 
one arm of 
the 
bridge, and its value is precisely determined from the amplitude and the phase 
of a 
nulling 
ac voltage applied to a standard capacitor (10-100fF) connected as the bridge's 
second 
arm.\cite{ashoori} Care was taken to set the measurement excitation level 
small enough
($<$100~$\mu$V) so that the data are observed to be amplitude independent.
In addition, a dc gate voltage $V_g$ is applied to the gate through a 
10M$\Omega$ thin-film resistor. It allows us to vary electron density $n_g$ 
under the 
gate, with larger positive $V_g$ yielding larger $n_g$.

	In a first experiment we determine the impedance between the gate and 
the ohmic 
contact to the 
2DEG as a function of $V_g$ in zero magnetic field.
We model this impedance as a resistance $R$ in series with a capacitance 
$C_g$ between 
the gate and the 
2DEG. A shunt parallel capacitance $C_{sh}$ is included in the model (see 
bottom inset 
to Fig.~1).
At low enough frequencies, $C_g$ becomes fully charged through $R$ during 
one 
cycle of the excitation.  In this case, the measured impedance is a capacitance 
of value 
$C_g+C_{sh}$. When the 2DEG under the gate is fully depleted, this area no 
longer charges, and the measured impedance is just $C_{sh}$.

	In magnetic field B the overall behavior becomes more complicated 
and 
interesting. First, we consider the case of integer Landau level filling factor in 
the bulk. 
Fig.~1 shows the dependence of the capacitance with the gate voltage at 
different 
frequencies for bulk filling factor $\nu_b=2$ in  sample~\#1. The measured 
capacitance is 
large at small negative $V_g$, and drops by an amount very nearly $C_g$ at 
small 
positive 
$V_g$. It returns to the larger value at greater positive $V_g$, albeit in a very 
nonmonotonic fashion. For higher frequencies, the capacitance minimum is 
widened, 
mostly from the high density (positive $V_g$) side.

	The overall picture can be simply described as follows. The bulk of the 
2DEG is 
effectively insulating at low temperature and, even for our lowest frequencies, 
charge 
cannot penetrate the bulk. Therefore, charge only appears in the compressible 
edge 
channel. Negative values of $V_g$ decrease the filling factor under the gate 
$\nu_g$ 
relative to the bulk value, and this conductive area merges with the edge 
channel (left 
inset 
to Fig.~1). Correspondingly, the measured capacitance is large in the whole 
frequency 
range. As $\nu_g$ is increased toward $\nu_b$, the entire area under the gate, 
except that 
associated with edge channels, is insulating and the capacitance drops (middle 
inset).  
When $\nu_g$  exceeds $\nu_b$, a conducting ``puddle'' of electrons in the 
next Landau 
level 
appears under the gate (right inset), but now this puddle is separated from the 
edge 
channel by an IS with filling factor $\nu_s=2$. Electrons must tunnel across 
the 
IS to charge the puddle.\cite{vaart}

The effective resistance $R$ of IS can be determined by 
fitting the frequency dependence of the capacitance to the model of the inset 
of Fig.~2. 
The 
capacitance has the form $C(f)=C_{high}+\delta 
C/[1+(2{\pi}R{\delta}Cf)^2]$, where 
$\delta C=C_{low}-C_{high}$ is the difference in the measured capacitance 
in the low 
and 
high frequency limits respectively. Fig.~1 thus demonstrates that conductance 
resonances 
give rise to the observed capacitance peaks.
	
	In Fig.~2, the capacitance of sample~\#2 at a frequency of 100 kHz is 
plotted as a 
function of $V_g$ for a range of B. Valleys clearly seen in the capacitance are 
associated with different integer $\nu_g$ ($\nu_g=$1,2,4). Let us 
follow the evolution of the valley around $\nu_g=1$. At low B 
(2.5~T$<B<$3.5~T), $\nu_b>1$, and minima are seen only at $\nu_g\simeq 
1$, when 
part of the 
area under the gate is insulating. In the intermediate 
B region (3.5~T$<B<$5~T) the bulk conducts poorly since $\nu_b\simeq 1$. 
The 
situation is
essentially the same as discussed for Fig.~1, and the mimima broaden on the 
$\nu_g>1$ 
side.
For the largest fields (5~T$<B<$6~T) shown in Fig.~2, the bulk is 
highly conducting, but now the conducting puddle under the gate with 
$\nu_g>1$ is 
separated from the bulk with $\nu_b<1$ and from the edge channel by an IS 
with 
$\nu_s=1$.
The other capacitance minima evolve in much the same fashion on the $V_g$-
B plane.

	In Fig.~2, the same resonance structure seen in Fig.~1 can be followed 
over a 
wide 
range of B. We believe that these resonances arise from tunneling between the 
edge and 
the puddle. They persist even for B at which the bulk acts as an insulator. 
Additionally, a 
steeper density gradient on the side of the puddle nearest the edge leads to a 
narrower IS 
there than on the bulk side\cite{chklov}.

We have found striking similarity in the overall resonance structure in all of 
the 
investigated samples. This semblance is particularly manifest for resonances 
in the strip 
with $\nu_s=2$. Fig.~3 shows the resonance positions detected on 
samples~\#1, \#2, and 
\#3. 
The positions are plotted as a function of  the gate voltage, measured from the 
sharp 
structure observed on the low density side of the capacitance minimum (see 
the inset to 
Fig.~3.) Our lowest frequency measurements indicate that the sharp structure 
is 
immediately 
followed by the appearance of the {\it compressible} puddle under the gate. 
As the inset 
illustrates, smaller samples display a wider capacitance minimum, and the 
average 
conductance per unit gate length is smaller. Our technique is thus most 
sensitive in 
different 
regions of gate voltage for each sample. Nonetheless, there is no doubt that the 
resonances 
displayed in each sample arise from similar origins. An average spacing of  
$\sim$40~mV 
is clearly observed in each case. The strength and exact positions of the 
resonances 
change 
as any individual sample is thermally cycled to room temperature, but the 
qualitative 
features of the resonant structure (average spacings and resonance widths) 
remain.

Higher temperatures smear the resonances both by increasing the non-resonant 
background conductance and by broadening the resonances. The temperature 
dependence 
of one of the resonances, characteristic for all observed resonances for 
$\nu_s=2$, is 
shown in 
the top inset to Fig.~4 for sample~\#3. The resonance broadens in an 
asymmetric fashion 
with increasing temperature, moving its center towards larger gate voltages, 
and the 
resonance width grows roughly linearly with the temperature.

The detailed frequency dependence of the capacitance is different at resonance 
positions and between resonances. Fig.~4 shows the capacitance measured at a 
resonance 
peak ($P_0$) and at nearby valley ($V_0$) as a function of frequency together 
with the 
best fit by simple $RC$-model (dashed lines) as described above. While on 
resonance the 
experimental points are fit well with characteristic frequency $f_c=$100kHz, 
the 
frequency 
dependence at the valley deviates significantly from the best fit with 
$f_c=5$kHz at high 
frequencies. Such deviation is characteristic for all valleys.

To illustrate this point better and to exclude possible systematic error over any 
given frequency range, we also plot in Fig.~4 data for a different peak ($P_2$) 
and valley 
($V_1$) chosen so that the corresponding characteristic frequencies are nearly 
the same. 
This deviation at the valleys indicates that a distributed resistance model 
should be used 
to 
fit the data rather than the simple $RC$ form described above. The simplest 
form for this 
model includes two resistances: one associated with tunneling through the IS 
($R_T$) and the other $R_X$ describing charge transfer through a transition 
region 
near the IS which separates it from the highly conducting interior of the 
puddle (see 
bottom inset to Fig.~4). Indeed, the shape of the experimental $C(f)$ curves in 
the valleys 
is indicative of charging of two regions. One small region charges quickly, 
creating a high 
frequency tail, and a second region which is about 10 times larger charges 
more slowly. 
The curve $V_0$ is fit based on this model (see Fig.~4). The parameters 
inferred from 
the 
fit are $R_T=7R_{peak}$ and $R_X=13R_{peak}$, where $R_{peak}$ 
($1.5\cdot 
10^7\Omega$) is 
the resistance at the resonance $P_0$.

Present models do not predict the observed resonant structure. According the 
electrostatic model of the edge developed by Chklovskii et al.\cite{chklov} the 
width of 
the IS is a monotonically decreasing function of the density gradient. As the 
filling factor 
in the ``puddle'' increases beyond the integer value, the width of the strip and 
consequently 
the tunneling resistance must decrease.

One might consider the presence of favorably situated impurities as the origin 
of 
the tunneling resonances. Increasing the electron density under the gate causes 
the IS to 
move toward the edge. Even if the energy of an impurity state differs strongly 
from the 
Fermi energy, tunneling may be facilitated due to the presence of a virtual 
state each time 
the strip crosses the impurity location. Hence, each individual resonance trace 
could be 
assigned to a single ``good'' impurity. We find however that this portrait is 
inadequate to 
describe our results.

The data of Fig.~3 show that samples with vastly different dimensions show 
nearly 
identical resonances, strongly implying a common origin. It is very unlikely 
that in each 
case only one impurity is important over similar gate bias intervals. Most 
probably, 
impurity assisted tunneling averaged over many sites causes the growth 
of the non-resonant background with gate length.

The observed asymmetric growth of the resonances with increased 
temperature 
also discounts the impurity-assisted tunneling model. For resonant tunneling 
the 
conductance peak is expected to be both narrower and, in contrast to the 
observed 
behavior, {\it higher} with decreased temperature. For non-resonant impurity 
assisted 
tunneling, conductance peaks should remain unchanged at temperatures below 
the energy 
difference between the Fermi energy and the impurity energy level, whereas 
all observed 
resonances continue to narrow at the lowest experimental temperatures.

In our opinion, to account for the similarity of the resonances in the different 
samples it is {\it necessary} to assume that the tunneling conditions are 
modified 
homogeneously along the length of the strip.  One possible origin of the 
resonances may 
be reconstruction of the electron gas at the edge due to the competition 
between the 
repulsive Coulomb interaction and the effectively attractive exchange 
interaction 
\cite{chamon,klein,ash2}. As of yet, there 
is no theory for these reconstructions in the ``soft edge'' regime of our samples.
Another idea centers on the fact that the electron density across the IS is not 
perfectly 
constant.
Nazarov\cite{nazarov} has predicted that such a density gradient may give rise 
to a
crystalline ordering of electrons in regions in and adjoining the IS.
We speculate that for gate biases off resonance, electrons in these regions are 
localized, 
impeding
charge transfer to the puddle and leading to the charging process
described in the bottom inset of Fig.~4. The positions of the resonances may 
correspond to particular density gradients at the location of the IS when one 
crystalline structure evolves into another, and regions adjoining the IS become 
more 
conductive. 

Two basic facts are known about the resonances. One, they are no broader and 
appear no more frequently in large samples than small ones. This suggests a 
phenomenon 
occurring ubiquitously along the edge. Two, their exact positions and heights 
differ upon 
thermal cycling of samples and are therefore sensitive to the microscopic 
configuration of the sample. We conclude with the notion that these two facts 
are 
compatible only if a phenomenon exists which forces large regions of the edge 
to behave 
as a single structure.

We gratefully acknowledge helpful discussions with D.B. Chklovskii, B.I. 
Halperin, M.A. 
Kastner, 
A.H. MacDonald, K. Matveev, S. Simon, and X.-G. Wen and the assistance of 
H.B. Chan 
and D. Berman.
This work is supported by 
the ONR, the Packard Foundation, and NSF DMR-9357226 and DMR-
9400415.

\begin{figure}

\epsfxsize=\linewidth
\epsfbox{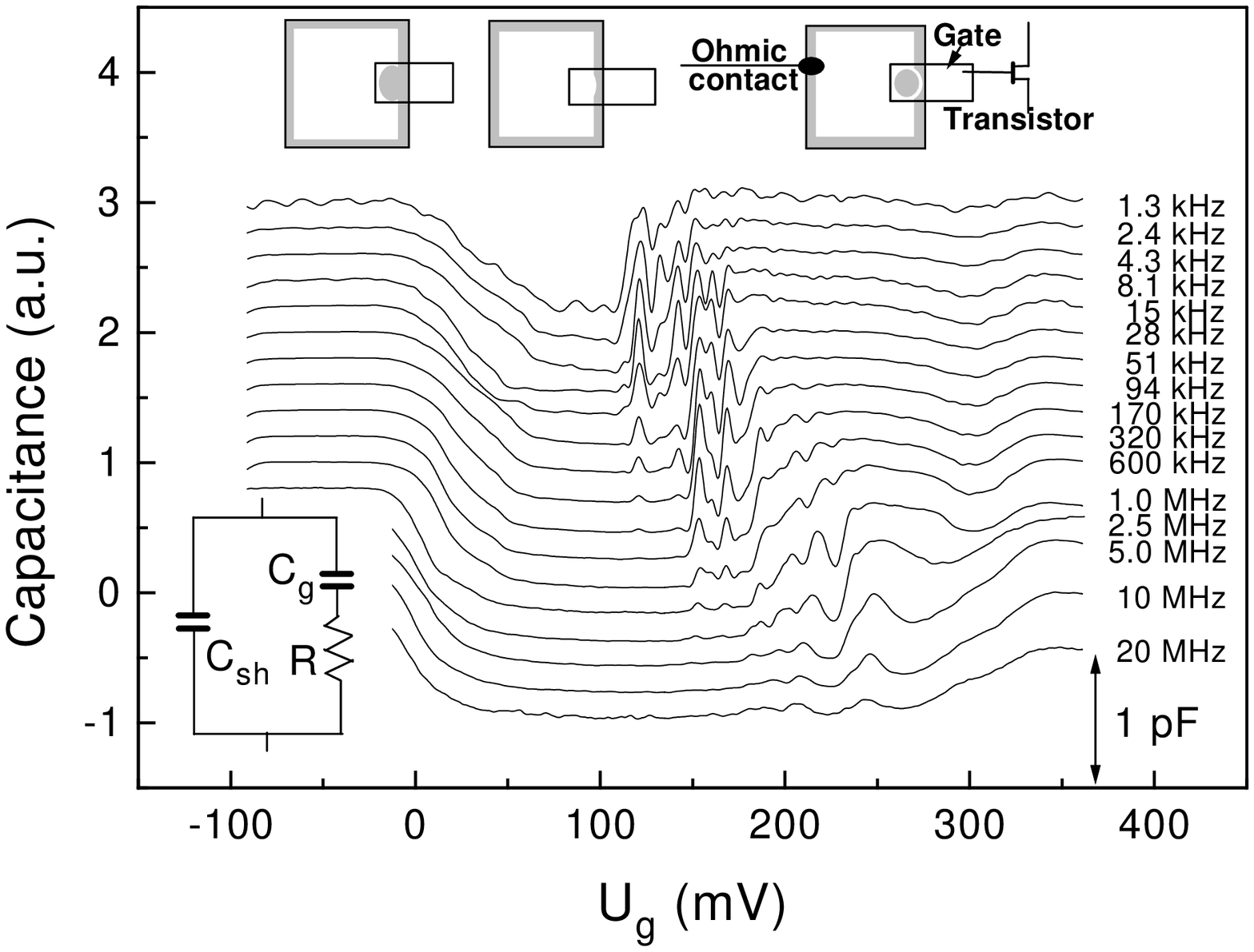}
\caption{Dependence of capacitance measured on  sample~\#1 (170~$\mu$m 
gate) 
with gate bias for different frequencies for B=2.1T (bulk filling factor 
$\nu_b=2$) at 
80 mK. Curves are shifted for clarity. Bottom inset: elementary electrical 
model of the 
sample. Top insets: region under the gate for different regimes of gate bias. 
White regions 
are incompressible, and shaded regions are compressible.}

\end{figure}

\begin{figure}

\epsfxsize=\linewidth
\epsfbox{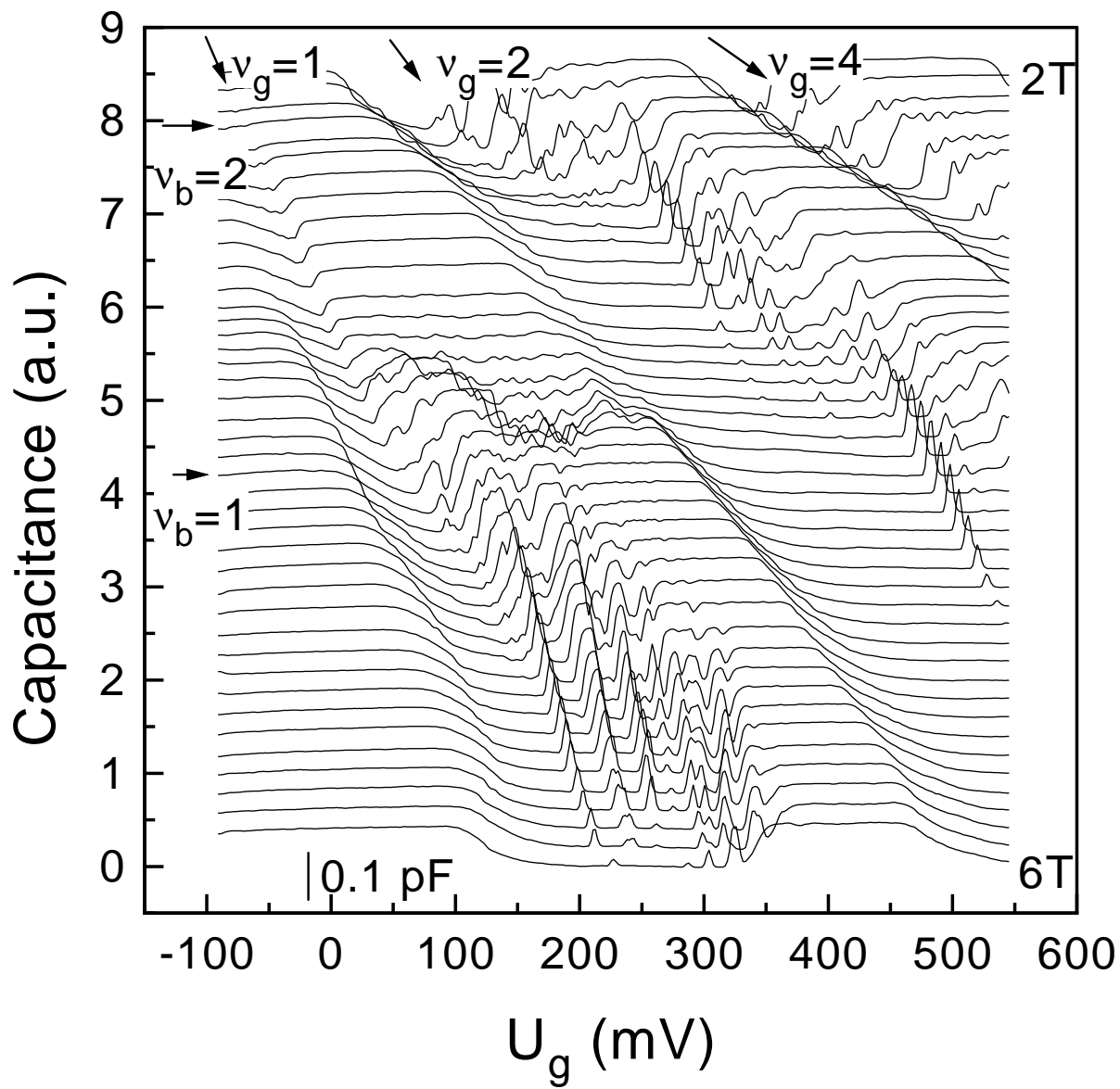}
\caption{Capacitance measured for  sample~\#2 at 100~kHz as a function of 
gate 
voltage shown for a region of magnetic field. T=80~mK. Curves are shifted 
for clarity. }

\end{figure}

\begin{figure}

\epsfxsize=\linewidth
\epsfbox{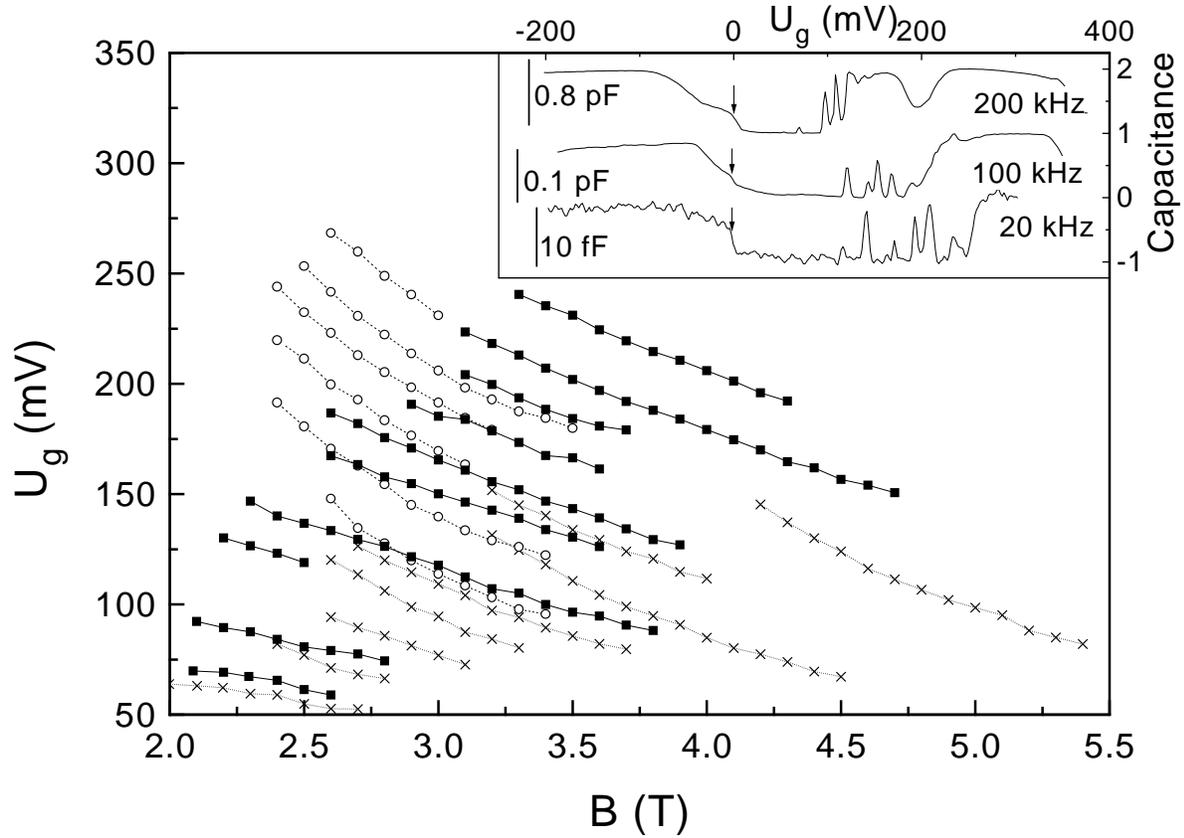}
\caption{Gate bias positions of the conductance resonances in the strip with 
$\nu_s=2$ observed for  sample~\#1 (170~$\mu$m gate, crosses), \#2 
(32~$\mu$m gate, 
squares) and \#3 (3~$\mu$m gate,  circles). Inset: normalized capacitance for 
these 
samples for B=3.0~T vs. the gate voltage.}

\end{figure}

\begin{figure}

\epsfxsize=\linewidth
\epsfbox{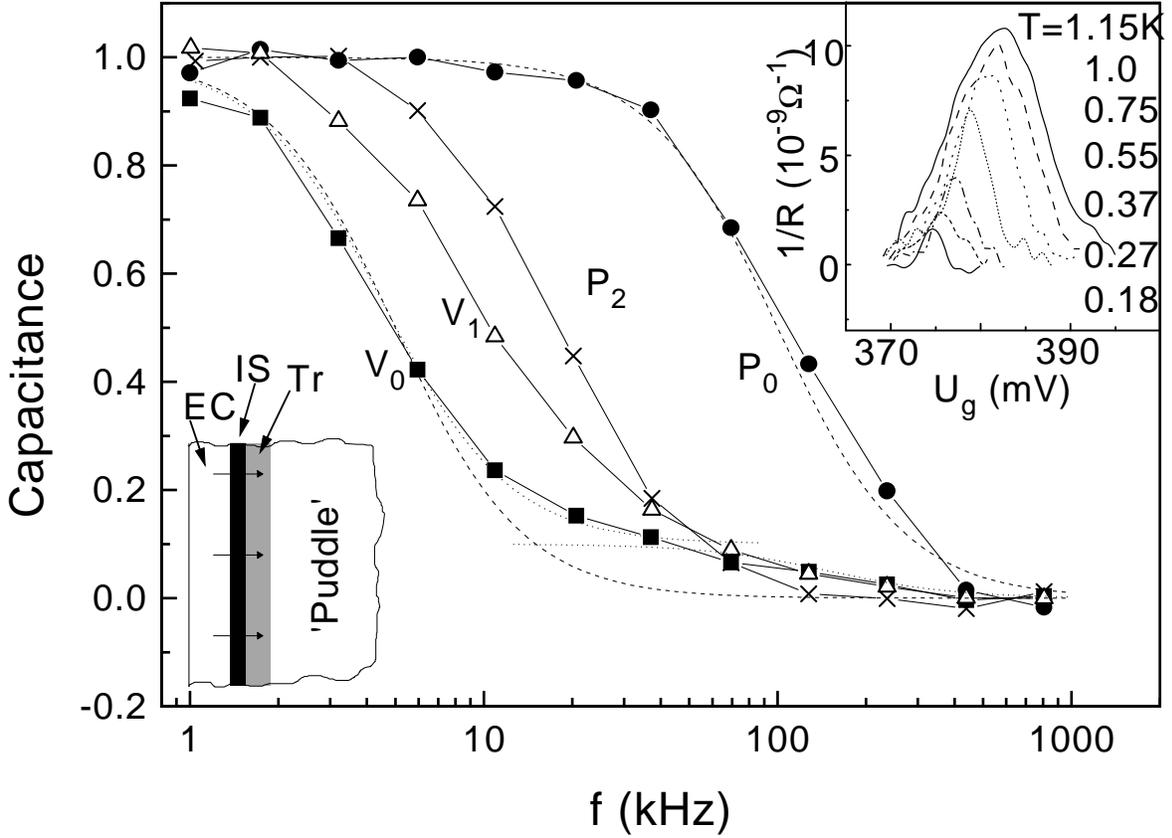}
\caption{Frequency dependences of the capacitance measured at the position 
of  
two conductance peaks ($P_0$ and $P_2$) and two valleys ($V_0$ and 
$V_1$) with 
$P_0$ 
and $V_0$ fitted (dashed lines) with the $RC$-model of Fig.~1 and the 
distributed-
resistance model (dotted line, for $V_0$ only). Data are for  sample~\#2 
(32~$\mu$m 
gate) 
at T=80mK.  Bottom inset: illustration of the off-resonance tunneling process 
from the 
edge channel (EC) through the incompressible strip (IS) into a poorly 
conducting 
transition region (Tr). Top inset: variation of a conductance resonance with 
temperature 
( sample~\#3, 3~$\mu$m gate, B=3.0~T).}

\end{figure}

\end{document}